\documentclass[11pt]{article}
\usepackage{graphicx}
\usepackage{amssymb}
\usepackage{amscd}
\usepackage{mathrsfs}
\usepackage{longtable,lscape}
\usepackage{amsthm}
\usepackage{amsfonts}
\usepackage{amsmath}
\usepackage{bbm}
\usepackage{float}
\usepackage{url}

\newcommand{\captionfonts}{\footnotesize}
\makeatletter  
\long\def\@makecaption#1#2{%
  \vskip\abovecaptionskip
  \sbox\@tempboxa{{\captionfonts #1: #2}}%
  \ifdim \wd\@tempboxa >\hsize
    {\captionfonts #1: #2\par}
  \else
    \hbox to\hsize{\hfil\box\@tempboxa\hfil}%
  \fi
  \vskip\belowcaptionskip}
\makeatother 
\begin{document}
\title{Spin and Wind Directions I: Identifying Entanglement \\ in Nature and Cognition}
\author{Diederik Aerts$^1$, Jonito Aerts Argu\"elles$^2$, Lester Beltran$^3$, Suzette Geriente$^4$, \\ Massimiliano Sassoli de Bianchi$^{1}$, Sandro Sozzo$^{5}$  and Tomas Veloz$^1$ \vspace{0.5 cm} \\ 
        \normalsize\itshape
        $^1$ Center Leo Apostel for Interdisciplinary Studies, 
         Brussels Free University \\ 
        \normalsize\itshape
         Krijgskundestraat 33, 1160 Brussels, Belgium \\
        \normalsize
        E-Mails: \url{diraerts@vub.ac.be,msassoli@vub.ac.be},\\\url{tveloz@gmail.com}
          \vspace{0.5 cm} \\ 
        \normalsize\itshape
        $^2$ KASK and Conservatory, \\
        \normalsize\itshape
         Jozef Kluyskensstraat 2, 9000 Ghent, Belgium
        \\
        \normalsize
        E-Mail: \url{jonitoarguelles@gmail.com}
	  \vspace{0.5 cm} \\ 
        \normalsize\itshape
        $^3$ 825-C Tayuman Street, \\
         \normalsize\itshape
        Tondo, Manila, The Philippines
         \\
        \normalsize
        E-Mail: \url{lestercc21@gmail.com}
	  \vspace{0.5 cm} \\ 
        \normalsize\itshape
        $^4$ Block 28 Lot 29 Phase III F1, \\ 
         \normalsize\itshape
        Kaunlaran Village, Caloocan City, The Philippines
         \\
        \normalsize
        E-Mail: \url{sgeriente83@yahoo.com}
	  \vspace{0.5 cm} \\ 
        \normalsize\itshape
        $^5$ School of Management and IQSCS, University of Leicester \\ 
        \normalsize\itshape
         University Road, LE1 7RH Leicester, United Kingdom \\
        \normalsize
        E-Mail: \url{ss831@le.ac.uk} 
       	\\
              }
\date{}
\maketitle
\begin{abstract}
\noindent
We present a cognitive psychology experiment where participants were asked to select pairs of spatial directions that they considered to be the best example of {\it Two Different Wind Directions}. Data are shown to violate the CHSH version of Bell's inequality with the same magnitude as in typical Bell-test experiments with entangled spins. Wind directions thus appear to be conceptual entities connected through meaning, in human cognition, in a similar way as spins appear to be entangled in experiments conducted in physics laboratories. This is the first part of a two-part article. In the second part \cite{ass2017} we present a symmetrized version of the same experiment for which we provide a quantum modeling of the collected data in Hilbert space.\end{abstract}

\section{Introduction\label{intro}}

Entanglement is one of the most characteristic manifestations of quantum structures and has been widely investigated both theoretically and experimentally. It is nowadays generally recognized that quantum entangled entities may exhibit non-local correlations that cannot be accounted for in a classical probabilistic framework, as formalized by Kolmogorov \cite{k1933}. Entanglement, however, is not a prerogative of micro-systems only. Its presence can be evidenced also in connected macro-systems \cite{Aerts1991, aabg2000, sdb2013a, sdb2013b} and in cognitive domains, when certain experiments are performed with human participants \cite{as2011, as2014}.

The scope of the present article is to present a cognitive psychology experiment where participants were interrogated about their preferences on wind directions, and show that there is a remarkable resemblance with typical coincidence experiments on spin entities in entangled states. More precisely, we will show that, when respondents jointly select two wind directions, probabilistically speaking they do so very much like how spin values along those same directions are selected by Stern-Gerlach apparatuses operating on spin bipartite entities in singlet states. This occurs because Bell's inequality (more precisely, its CHSH version) is violated in both domains with equivalent numerical magnitude, thus allowing to conclude about the detection of entanglement in cognitive experiments, similarly to how it is typically detected in physics experiments. 

The article is organized as follows. In Sec.~\ref{quantumentanglement}, we describe the situation of a typical (EPR-Bohm like) `Bell-test experiment' in physics, indicating what the quantum mechanical prediction is, as far as the violation of Bell's inequality is concerned. Then, in Sec.~\ref{theoreticalexperiment}, we describe our experiment with the conceptual {\it Two Different Wind Directions} entity, and its results, highlighting how Bell's inequality is violated in a way that is very similar to as it is violated in physics. Finally, in Sec.~\ref{conclusion}, we provide some final remarks.

\section{Entanglement in a typical CHSH (two-channel) experiment\label{quantumentanglement}}

The seminal studies of John Bell on the foundations of quantum theory showed that a measurement on a quantum entity does not generally reveal a pre-existing value of the measured quantity, which is instead actualized by the measurement context (`quantum contextuality'), and that such process of `actualization of potential properties' also occurs when the measurement context is formed by parts that are separated by large spatial distances. In other words, quantum contextuality also holds at a distance (`quantum non-locality'). The latter effect is determined by what physicists call, more specifically, `quantum entanglement'. Effects due to the presence of quantum entanglement were firstly identified in the seventies of the foregoing century by experiments at the time not yet completely convincing \cite{freedman1972,holt1973,fry1976,lamehi1976}, which culminated in 1982 in the major photon correlation experiment performed by the team of Alain Aspect in Paris \cite{aspect1982a, aspect1983}. More recent experiments (see, e.g., \cite{aspect1082b,tittel1998,weihs1998,genovese2005,vienna2013,urbana2013,hensen-etal2015}) have eventually `closed all loops', confirming that quantum theory properly describes the situation, thus certifying the `reality of quantum entanglement as a phenomenon'.

\begin{figure}[htbp]
\begin{center}
\includegraphics[width=17cm]{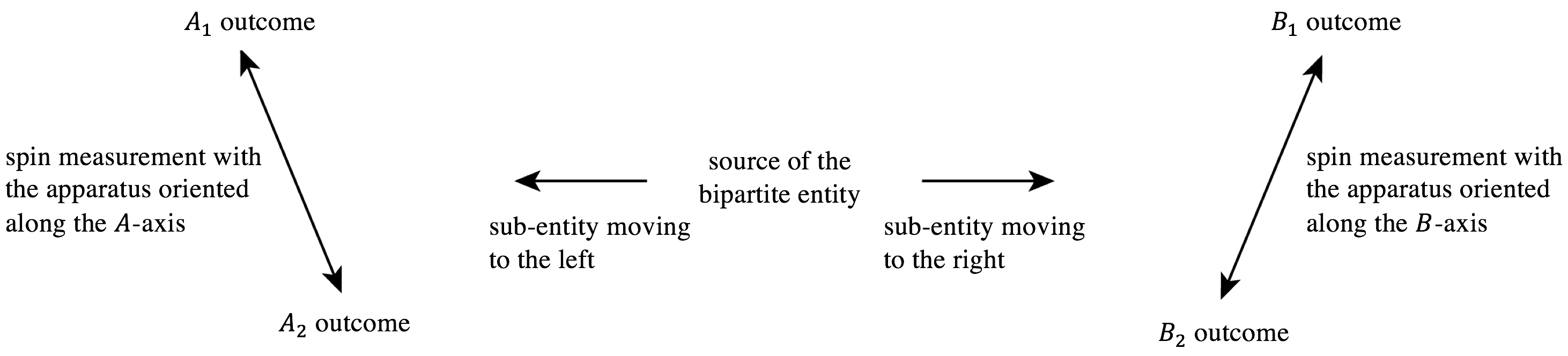}
\caption{A schematic representation of a typical Bell test experimental setting in physics.} 
\label{EPRSituation}
\end{center}
\end{figure}

Let us describe the experimental setting of a typical Bell-test experiment, and how it is modeled by quantum theory. A source prepares a bipartite compound entity in a state characterized by an overall spin value equal to zero. Also, the state of the bipartite entity is such that the two sub-entities forming it, which are assumed to be of spin-${1\over 2}$, fly apart in opposite spatial directions. Measuring apparatuses are in regions of space that are located symmetrically with respect to the source, along the direction of propagation of the two sub-entities, allowing for coincident measurements of their spin values, along given axes. If the spin of the sub-entity moving to the left is measured along the $A$-axis, there will be two possible outcomes: a `spin up' outcome and a `spin down' outcome, which will be denoted $A_1$ and $A_2$, respectively. Similarly, if the spin of the sub-entity moving to the right is measured along the $B$-axis, the `spin up' and a `spin down' outcomes will be denoted $B_1$ and $B_2$, respectively (a schematic representation of the experiment is presented in Fig.\ref{EPRSituation}).

The experiments that have been performed show that the outcomes of the joint spin measurements on the two sub-entities correlate in a very special way, in close accordance with quantum mechanical  predictions, thus making a convincing case for the hypothesis that quantum theory provides a faithful modeling of the situation, i.e., that the bipartite entity is in an entangled pre-measurement state. More precisely, the values of the probabilities characterizing these correlations, predicted by quantum theory, are as follows (as reported by manuals of quantum mechanics). If $\alpha$ is the angle between the $A_1$ and $B_1$ (respectively, $A_2$ and $B_2$, $A_1$ and $B_2$, $A_2$ and $B_1$) directions, then the probability of finding the $A_1$ and $B_1$ (respectively, $A_2$ and $B_2$, $A_1$ and $B_2$, $A_2$ and $B_1$) outcomes is: ${1 \over 2}\sin^2{\alpha \over 2}$. 

In concrete experiments, an angle of $45^\circ$ is usually chosen between the $A_1$ and $B_1$ directions and the $A_2$ and $B_2$ directions, hence there is also an angle of $180^\circ-45^\circ=135^\circ$ between the $A_1$ and $B_2$ directions and the $A_2$ and $B_1$ directions, as described in Fig.~\ref{EPRSituation} (the reason for this particular choice is that it produces correlations for which the presence of entanglement becomes most visible). This means that the probability for the $A_1$ and $B_1$ directions to correlate is $p(A_1,B_1)= {1 \over 2}\sin^2{45^\circ \over 2}={1\over 8}(2-\sqrt{2})\approx 0.0732$, and same for $p(A_2,B_2)$. Also, the probability for the $A_1$ and $B_2$ directions to correlate is $p(A_1,B_2)={1 \over 2}\sin^2{135^\circ \over 2}={1\over 8}(2+\sqrt{2})\approx 0.4268$, and same for $p(A_2,B_1)$. Note that $p(A_1,B_1)+p(A_2,B_2)+p(A_1,B_2)+p(A_2,B_1)={2\over 8}(2-\sqrt{2})+{2\over 8}(2+\sqrt{2})=1$, which means that the measurements give rise with certainty to one of the four possible correlations (this is the idealized situation of detectors having $100 \%$ efficiency).

To explain how entanglement can be deduced from the observed correlations, one needs to introduce Bell's inequality \cite{bell1964}, and more specifically its variant called CHSH inequality, due to Clauser, Horne, Shimony and Holt \cite{clauser1969}. But first, let us explain which aspect of our physical reality is tested by the latter. If we consider the experimental situation presented in Fig.~\ref{EPRSituation}, there is an obvious analogy with ordinary reality that comes to mind, namely the situation of the explosion of a material object into two fragments, one flying to the left and the other to the right. Obviously, these two fragments of the initial material object will manifest different types of correlations. To name an obvious one, if the object has a color, the two fragments will have the same color. The weights of the two fragments will also be perfectly correlated, as their sum must be equal to the weight of the unexploded object. The same is true for their momenta, which for instance will have to sum to zero in case before the explosion the object was at rest. The distances of the fragments from where the explosion took place, at a given moment, are also correlated, in ways that depend on their masses and momenta. Equally so, if rotation is involved, there will be a correlation of the angular momenta, which will be opposite in direction in case the material object had no initial rotational movement. 

If we assume that some indeterminism is involved, i.e., that we lack knowledge about the exact initial state of the material object, then the previously mentioned physical quantities associated with the two flying apart fragments are only describable in probabilistic terms, which means that also correlations will be described probabilistically. However, not all combinations of probabilities describing the correlations can make their appearance in situations like the one of the exploding object, and it is precisely this fact, that not all combinations are possible, which is the main content of Bell and CHSH inequalities. 

Bell chose to consider expectation values instead of correlation probabilities, which is what we will also do now. When jointly performing measurement $A$ on the left entity and measurement $B$ on the right entity (the situation considered in Fig.~\ref{EPRSituation}), we will attribute the value $+1$ if the two spin outcomes are both `up' or both `down' (outcomes $A_1$ and $B_1$, or $A_2$ and $B_2$), and the value $-1$ if one of them is `up' and the other one `down' (outcomes $A_1$ and $B_2$, or $A_2$ and $B_1$). The expectation value for the joint measurement is then given by: 
\begin{equation}
E(A,B)=p(A_1,B_1)-p(A_1,B_2)-p(A_2,B_1)+p(A_2,B_2).
\end{equation}
Hence, if the angle between $A_1$ and $B_1$ is $45^\circ$, the quantum model predicts the expectation value: 
\begin{equation}
E(A,B)={2\over 8}(2+\sqrt{2}) - {2\over 8}(2-\sqrt{2}) = -{1\over\sqrt{2}}\approx -0.7071.
\end{equation}
To formulate the CHSH inequality, we need to consider, in addition to the joint measurement of $A$ and $B$ illustrated in Fig.~\ref{EPRSituation} (which we will simply denote $AB$), three other joint measurements, $AB'$, $A'B$ and $A'B'$, where $A'$ is another measurement (different from $A$) that can be performed on the left entity and $B'$ is another measurement (different from $B$) that can be performed on the right entity. Expectation values $E(A,B')$, $E(A',B)$ and $E(A',B')$ can then also be associated to these other three joint measurements, and the CHSH inequality is:
\begin{equation}\label{chsh}
|S|\le 2, \quad S\equiv E(A,B)-E(A,B')+E(A',B)+E(A',B').
\end{equation}

Without entering into a specific discussion of the notion of `local realism', assumed by Bell to state and prove the `no-go theorem' associated with the above and similar inequalities, it is sufficient for us to observe that Bell's characterization is such that no bipartite system of the `material object exploded into two fragments' kind can ever violate (\ref{chsh}). To give an example, consider the situation where $A$ ($B$) corresponds to the measurement of the weight of the left (right) fragment, with the outcome $A_1$ ($B_1$) corresponding to the situation where the weight is more than half the weight of the initial object, and $A_2$ ($B_2$) to the situation where it is less. Also, assuming that the initial object is red, $A'$ ($B'$) is taken to be the measurement of the color of the left (right) fragment, with the outcome $A_1$ ($B_1$) corresponding to the situation where the red color is obtained, and $A_2$ ($B_2$) to the situation where a color different than red is obtained. Clearly, we have the probabilities $p(A'_1,B'_1) = 1$ and $p(A_1,B'_2)= p(A'_2,B_1)= p(A'_2,B'_2)=0$, since both fragments are necessarily red. Also, since the two fragments cannot both weigh more or less than half the weight of the initial object, we have $p(A_1,B_1)=p(A_2,B_2)=0$. If $p$ denotes the probability that the initial state of the object is such that the explosion will cause the right fragment to be heavier than the left one, we have: $p(A_2,B_1)=p$ and $p(A_1,B_2)=1-p$. Also, $p(A'_1,B_1)= p$ and $p(A_1,B'_1)= 1-p$, and similarly $p(A'_1,B_2)= 1-p$ and $p(A_2,B'_1)= p$. Replacing these probabilities in the expectation value formulae, we obtain $E(A,B)=-1$, $E(A,B')=1-2p$, $E(A',B)=-1+2p$ and $E(A',B')=1$, so that $S=-1-1+2p-1+2p+1=-2(1-2p)$. Thus, $|S|\le 2|1-2p|\le 2$, in accordance with (\ref{chsh}).

To show that the CHSH inequality is instead violated in spin coincidence measurements, we have to also define the two additional measurements $A'$ and $B'$ in that context. The latter is chosen to be a spin measurement with the apparatus oriented in such a way that there is an angle of $45^\circ$ between directions $A_1$ and $B'_2$, and directions $A_2$ and $B'_1$, hence also an angle of $135^\circ$ between directions $A_1$ and $B'_1$, and directions $A_2$ and $B'_2$. On the other hand, $A'$ is taken to be such that there is angle of $45^\circ$ between directions $A'_1$ and $B'_1$, and directions $A'_2$ and $B'_2$, hence an angle of $135^\circ$ between directions $A'_1$ and $B'_2$, and directions $A'_2$ and $B'_1$. This means that quantum theory predicts the probabilities $p(A_1,B'_2)=p(A_2,B'_1)=p(A'_1,B_1)=p(A'_2,B_2)=p(A'_1,B'_1)=p(A'_2,B'_2)={1\over 8}(2-\sqrt{2})\approx 0.0732$ and $p(A_1,B'_1)=p(A_2,B'_2)=p(A'_1,B_2)=p(A'_2,B_1)=p(A'_1,B'_2)=p(A'_2,B'_1)={1\over 8}(2+\sqrt{2})\approx 0.4268$. If we attribute again the value $+1$ for the 
`both spin up/down outcomes' and $-1$ to the `one spin up and one spin down outcomes', the expectation value of the four joint measurement $AB$, $AB'$, $A'B$ and $A'B'$ are: 
\begin{eqnarray}
&&E(A,B)=p(A_1,B_1)-p(A_1,B_2)-p(A_2,B_1)+p(A_2,B'_2)=-1/\sqrt{2}\approx -0.7071\nonumber\\
&&E(A,B')=p(A_1,B'_1)-p(A_1,B'_2)-p(A_2,B'_1)+p(A_2,B'_2)=1/\sqrt{2}\approx 0.7071\nonumber\\
&&E(A',B)=p(A'_1,B_1)-p(A'_1,B_2)-p(A'_2,B_1)+p(A'_2,B_2)=-1/\sqrt{2}\approx -0.7071\nonumber\\
&&E(A',B')=p(A'_1,B'_1)-p(A'_1,B'_2)-p(A'_2,B'_1)+p(A'_2,B'_2)=-1/\sqrt{2}\approx -0.7071,
\end{eqnarray}
so that we obtain the following violationof (\ref{chsh}): 
\begin{equation}
|S|=|E(A,B)-E(A,B')+E(A',B)+E(A',B')| = 2\sqrt{2}\approx 2.8284.
\end{equation}

In other words, the quantum mechanical prediction, nowadays confirmed by a considerable amount of experimental data \cite{aspect1082b,tittel1998,weihs1998,genovese2005,vienna2013,urbana2013,hensen-etal2015}, is that the CHSH inequality is violated when joint measurements are performed on bipartite systems, like spin systems, prepared in an entangled state. This is a situation that cannot be properly modeled by a classical (Kolmogorovian) probability theory. However, this doesn't mean that the CHSH inequality cannot be violated also by classical macroscopic systems. For this, it is sufficient that the two components remain connected in some way, so that the left and right measurements can influence each other's states and therefore outcomes \cite{ass2017,aabg2000,sdb2013a,sdb2013b,AertsSassoli2016}. This is what can be expected to happen also with entangled quantum entities, although their connection remains in this case hidden, i.e., appears to be a `non-spatial connection', hence the strangeness of the quantum entanglement phenomenon, famously referred to by Einstein as ``spooky action at a distance.''

\section{Entangled wind directions in human cognition \label{theoreticalexperiment}}

In this section, we consider joint measurements performed on a conceptual entity formed by the combination of two concepts, to highlight the presence of entanglement in human cognition and its similarity with entanglement of micro-physical bipartite entities. More precisely, we consider the following combination of concepts: {\it Two Different Wind Directions}. If analyzed from the perspective of its meaning, it is a combination of the following two conceptual elements: {\it One Wind Direction} and {\it Another Wind Direction}. However, in the English language, this combination is usually expressed by the sentence: {\it Two Different Wind Directions}. Our investigation of this conceptual combination is to consider measurements that can be performed on the two composing concepts, and analyze the statistics of outcomes associated with the combinations of these measurements. Indeed, it is in this statistics that traces can be be found of the presence of a quantum structure of the entanglement kind.

\begin{figure}[htbp]
\begin{center}
\includegraphics[width=17cm]{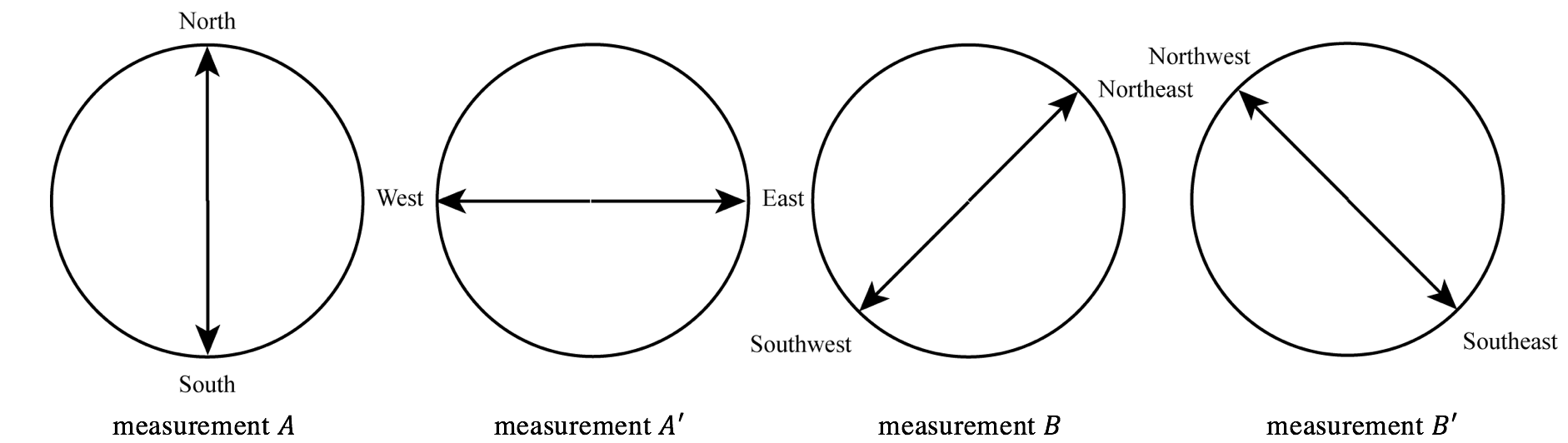}
\caption{A graphical representation of the two outcomes for the four measurements $A$, $A'$, $B$ and $B'$.}
\label{default}
\end{center}
\end{figure}

To explain how measurements are introduced and analyzed, let us first consider the single conceptual element {\it One Wind Direction}, which together with the element {\it Another Wind Direction} is part of the combination {\it Two Different Wind Directions}. A typical measurement is to ask a human participant to choose one among two possible wind directions. For example, either the directions {\it North} or the direction {\it South}, giving rise to a two-outcome measurement that we will denote $A$. To perform a typical Bell-test experiment, we need to define three additional measurements: $A'$, $B$ and $B'$. Measurement $A'$ is also considered to apply to the conceptual entity {\it One Wind Direction}, and consists in choosing among the two wind directions {\it East} and {\it West}. Measurement $B$ and $B'$ are instead considered to apply to the conceptual entity {\it Another Wind Direction}, and consists in asking a human participant to choose between the directions {\it Northeast} and {\it Southwest}, and {\it Southeast} and {\it Northwest}, respectively (see Fig.~\ref{default}).

\begin{figure}[htbp]
\begin{center}
\includegraphics[width=16cm]{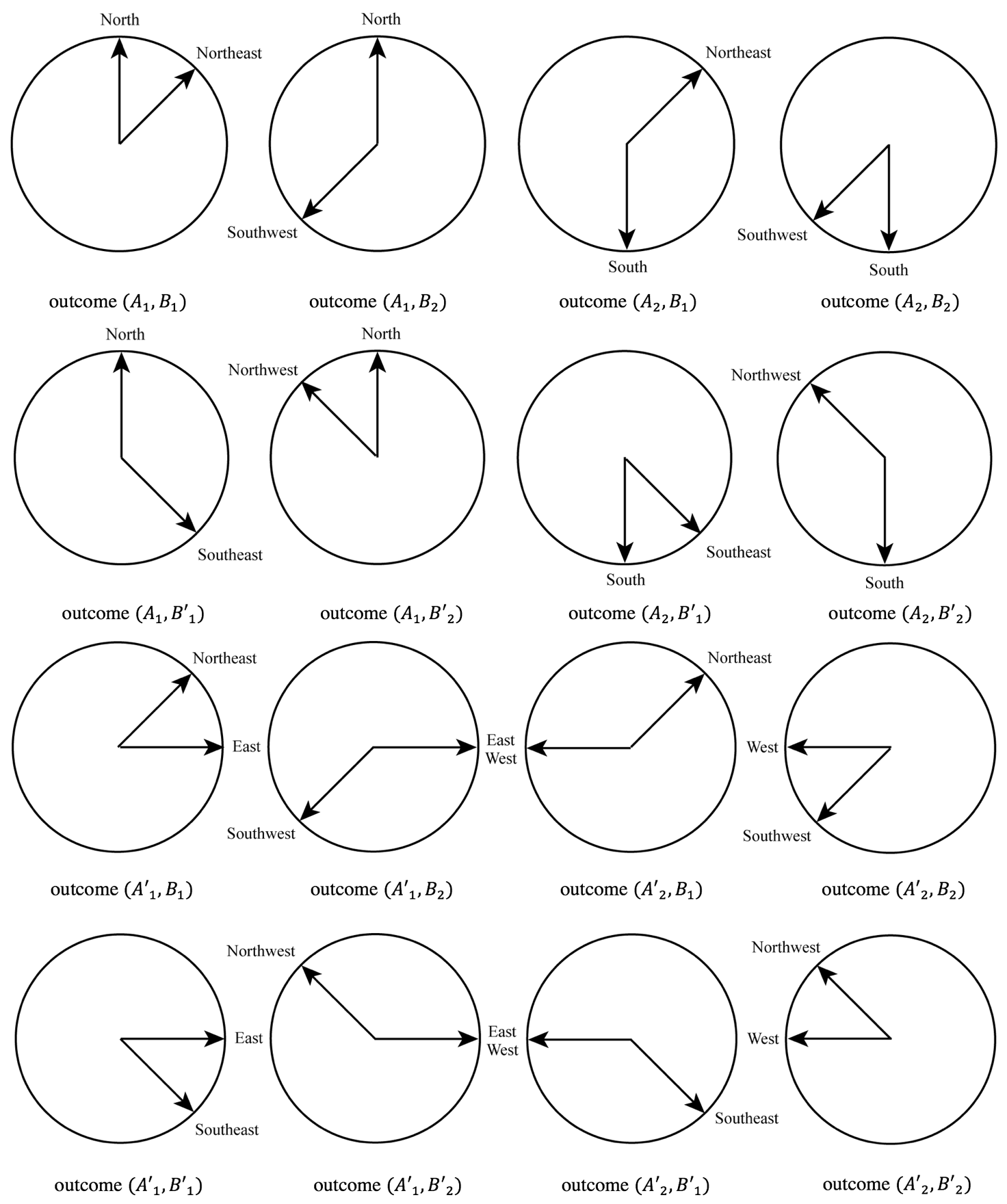}
\caption{A graphical representation of the four possible outcomes of the joint measurements $AB$, $AB'$, $A'B$ and $A'B'$.} 
\label{default2}
\end{center}
\end{figure}

To be in the situation where the CHSH inequality can be tested, we need then to consider the joint measurements that can be defined by combining the above four measurements. More specifically, we denote $AB$ the joint measurement of $A$ combined with $B$, $AB'$ the joint measurement of $A$ combined with $B'$, $A'B$ the joint measurement of $A'$ combined with $B$, and $A'B'$ the joint measurement of $A'$ combined with $B'$. These are measurements that are now to be performed on the combined concept {\it Two Different Wind Directions}. More precisely, joint measurement $AB$ consists of a human subject choosing one among four possible outcomes that are combinations of the outcomes of measurements $A$ and $B$. Hence, the possible outcomes of $AB$ are {\it North} and {\it Northeast} [outcome $(A_1,B_1)$], {\it North} and {\it Southwest} [outcome $(A_1,B_2)$], {\it South} and {\it Northeast} [outcome $(A_2,B_1)$], and {\it South} and {\it Southwest} [outcome $(A_2,B_2)$]. Similarly, joint measurement $AB'$ consists of a human subject choosing one among four possible outcomes that are combinations of the outcomes of measurements $A$ and $B'$. Hence, the possible outcomes of $AB'$ are {\it North} and {\it Southeast} [outcome $(A_1,B'_1)$], {\it North} and {\it Northwest} [outcome $(A_1,B'_2)$], {\it South} and {\it Southeast} [outcome $(A_2,B'_1)$], and {\it South} and {\it Northwest} [outcome $(A_2,B'_2)$]. Also, joint measurement $A'B$ consists of a human subject choosing one among four possible outcomes that are combinations of the outcomes of measurements $A'$ and $B$. Hence, the possible outcomes of $A'B$ are {\it East} and {\it Northeast} [outcome $(A'_1,B_1)$], {\it East} and {\it Southwest} [outcome $(A'_1,B_2)$], {\it West} and {\it Northeast} [outcome $(A'_2,B_1)$], and {\it West} and {\it Southwest} [outcome $(A'_2,B_2)$]. Finally, joint measurement $A'B'$ consists of a human subject choosing one among four possible outcomes that are combinations of the outcomes of measurements $A'$ and measurement $B'$. Hence, the possible outcomes of $A'B'$ are {\it East} and {\it Southeast} [outcome $(A'_1,B'_1)$], {\it East} and {\it Northwest} [outcome $(A'_1,B'_2)$], {\it West} and {\it Southeast} [outcome $(A'_2,B'_1)$], and {\it West} and {\it Northwest} [outcome $(A'_2,B'_2)$]; see Fig.~\ref{default2}.

Having described the four joint measurements $AB$, $AB'$, $A'B$ and $A'B'$, let us now explain how they were performed in practice. Concerning `participants and design', we asked 85 persons, chosen at random among colleagues and friends, to fill in a questionnaire with closed-ended questions. The experimental design was a `repeated measures', or `within subjects' design, which means that all participants were subject to the same questions and experimental conditions. Concerning `procedure and materials', the questionnaire consisted in four sequential tests, where each test is a question with four possible answers, with the possibility to only pick one answer. The different answers were accompanied by their graphical description. For instance, the answer ``North and Northeast'' was associated with the first drawing of Fig.~\ref{default2}, the answer ``North and Southwest'' with the second drawing of Fig.~\ref{default2}, and so on. More precisely, the first test (denoted test A) corresponded to the joint measurement $AB$, the second test (denoted test B) to the joint measurement $AB'$, the third test (denoted test C) to the joint measurement $A'B$, and the last test (denoted test D), to the joint measurement $A'B'$. Before executing these four tests, an introductory text was also presented to the participants, which is the following: 

``This study has to do with what we have in mind when we use words that refer to categories, and more specifically `how we think about examples of categories'. Let us illustrate what we mean. Consider the category `fruit'. Then `orange' and `strawberry' are two examples of this category, but also `fig' or `olive' are examples of the same category. In each test of the questionnaire you will be asked to pick one of the examples of a set of given examples for a specific category. And we would like you to pick that example that you find `the best example' of the category. In case there are more than one example which you find the best example, pick then the one you prefer anyhow in some way. In case there are two examples which you both find the best, and hence hesitate which ones to take, just take then the one you slightly prefer, however slight the preference might be. In case you really have no preference, you are allowed to pick at random, and even use a coin to do so. It is mandatory that you always `pick one and only one example'. So, one of the tests could be that the category `fruit' is given, and you are asked to pick one of the examples `orange', `strawberry', `fig' or `olive', and hence choose amongst these the one you find `the best example'. What is also very important, let all aspects of yourself play a role in the choice you make, ratio, but also imagination, feeling, emotion, and whatever. Hence, suppose you feel like choosing a specific example as the best, but your ratio argues, eventually after analysis, that it is another one which is objectively the best, and then you are allowed (not obliged however) to stay with your feeling, and not follow the analysis. There are four tests, test A, test B, test C, and test D, explained on the next four pages, and hence for each of them one choice to be made.''

In addition to that, each single test was preceded by the following text: ``Consider the sentence {\it Two Different Wind Directions} and choose which of the following examples is the best example of {\it Two Different Wind Directions}. In case you find more than one best example, pick the one you find really the best, whatever aspect of yourself tells you that this is the one you prefer. You must pick one and only one, even in case you cannot decide after a while. If this would happen, namely that you keep hesitating, choose then one at random (you may use a coin for this). Put a cross behind the one you choose.''

For each one of the four tests, that is, for each one of the four joint measurements $AB$, $AB'$, $A'B$ and $A'B'$, we then calculated from the collected data the relative frequencies of the different outcomes (which in the large number limit can be interpreted as probabilities). The obtained experimental values are: 
\begin{eqnarray}
&&p(A_1,B_1)=0.13,\quad p(A_1,B_2)=0.55,\quad p(A_2,B_1)= 0.25,\quad p(A_2,B_2)=0.07 \nonumber\\
&&p(A_1,B'_1)=0.47,\quad p(A_1,B'_2)= 0.12,\quad p(A_2,B'_1)= 0.06,\quad p(A_2,B'_2)=0.35 \nonumber\\
&&p(A'_1,B_1)=0.13,\quad p(A'_1,B_2)=0.38,\quad p(A'_2,B_1)= 0.42,\quad p(A'_2,B_2)=0.07 \nonumber\\
&&p(A'_1,B'_1)= 0.09,\quad p(A'_1,B'_2)= 0.44,\quad p(A'_2,B'_1)=0.38,\quad p(A'_2,B'_2)=0.09,
\end{eqnarray}
so that the corresponding expectation values are: 
\begin{eqnarray}
&&E(A,B)=p(A_1,B_1)-p(A_1,B_2)-p(A_2,B_1)+p(A_2,B_2)=-0.6 \nonumber\\
&&E(A,B')=p(A_1,B'_1)-p(A_1,B'_2)-p(A_2,B'_1)+p(A_2,B'_2)=0.65 \nonumber\\
&&E(A',B)=p(A'_1,B_1)-p(A'_1,B_2)-p(A'_2,B_1)+p(A'_2,B_2)=-0.6 \nonumber\\
&&E(A',B')=p(A'_1,B'_1)-p(A'_1,B'_2)-p(A'_2,B'_1)+p(A'_2,B'_2)=\nonumber-0.62.
\end{eqnarray}
It follows that we find the following violation of the CHSH inequality: 
\begin{equation} \label{CHSHreal}
|S|=|E(A,B)-E(A,B')+E(A',B)+E(A',B')| =2.47.
\end{equation}

Equation (\ref{CHSHreal})  shows a striking similarity with the values of the violation obtained in typical physics experiments designed to detect entanglement and non-locality in spin-like coincidence measurements on pairs of quantum entities (electrons, ions, photons). For example, according to \cite{genovese2005}, Aspect et al. (1982) found $|S|=2.697\pm 0.015$, Tittel et al. (1998) found $|S|=2.38\pm 0.16$, Weihs et al. (1998) found $|S|=2.73\pm 0.02$, Aspelmeyer et al. (2003) found $|S|=2.41\pm 0.10$, Pittman \& Franson (2003) found $|S|=2.44\pm 0.13$, and Peng et al. (2004) found $|S|=2.45\pm 0.09$. Also, Hensen et al. (2015) found $|S|=2.42\pm 0.20$.

Note that we performed a statistical analysis of our experimental data, to test whether the observed deviation from the value $|S|=2$ was only due to chance. To this end, we computed a one tail one sample t-test for means of the experimental values of $|S|$ in Equation (\ref{CHSHreal}), against the constant value $2$, finding a $p$-value $p(df=84)=0.05$. This is a borderline result with respect to the rejection of the null hypothesis that the two means are equal in the t-test, but most probably this is only due to the reduced size of the statistical sample.

\section{Concluding remarks\label{conclusion}}

The results presented in the previous section show that conceptual entities can violate Bell's inequality (here the CHSH version of it) and that the magnitude of the violation is the same as that obtained in physics experiments with entangled pairs of spin-${1\over 2}$ entities (or similar quantum entities, like entangled photons). This closeness of values between cognitive and physics experiments is particularly striking because `joint spin measurements' and `joint wind measurements' are both about actualizing potential spatial directions. This means that people appear to actualize spatial directions (here in association with winds) in a way that is remarkably similar, statistically speaking, to how spin directions (more precisely, `up' and `down' directions, along given axes) are actualized by physical apparatuses, like Stern-Gerlach apparatuses. 

We mentioned in Sec.~\ref{quantumentanglement} that quantum entanglement can be understood as being the result of a connection between the two sub-entities forming the bipartite system. It is worth mentioning that this interpretation is supported by the `extended Bloch representation of quantum mechanics', where entangled states of bipartite systems can be written in a way that the two sub-systems always remain in well-defined states, their entanglement being instead associated with a third `element of reality', corresponding to the emergence of a non-spatial connection between them \cite{AertsSassoli2016} (which cannot be deduced from the states of the sub-entities, in accordance with the principle that the whole is greater than the sum of its parts). Coincidence measurements can therefore be understood as processes during which the symmetry of such connection between the sub-entities is broken, bringing them in a condition of (temporary) separation. This `symmetry breaking' (the actual breaking the symmetry of the potential) is a process that genuinely creates correlations that were not present before the measurement (called `correlations of the second kind' \cite{Aerts1991}), and it is precisely such process of `creation of correlations', as opposed to a process of mere `discovery of already existing correlations' (called `correlations of the first kind', as in the situation of the exploded object described in Sec.~\ref{quantumentanglement}) that is responsible for the violation of the CHSH inequality. 

When dealing with entanglement in human cognition, the nature of the non-spatial connection responsible for the violation of the CHSH inequality can be understood as a `connection through meaning'. Indeed, before asking the participants to choose a pair of specific directions, an abstract meaning-connection undoubtedly exists between the two conceptual wind directions, expressing all their possible concrete actualizations (instantiations) that are meaningful for us humans, in view of our experience with Euclidean space and its directions. And when pairs of specific directions are actualized, the symmetry of such abstract meaning-connection is broken, creating in this way correlations.

We conclude by evoking an important aspect in the violation of Bell's and CHSH inequalities: the preservation of so-called `marginal law', which is typically assumed in their derivation \cite{f1982}.\footnote{The marginal law, in the ambit of Bell-test experiments, expresses the classical Kolmogorovian requirement that the following eight equalities must be fulfilled: $\sum_{j=1}^2p(A_i,B_j)= \sum_{j=1}^2p(A_i,B'_j)$, $i=1,2$; $\sum_{i=1}^2p(A_i,B_j)= \sum_{i=1}^2p(A'_i,B_j)$, $j=1,2$; $\sum_{j=1}^2p(A'_i,B_j)= \sum_{j=1}^2p(A'_i,B'_j)$, $i=1,2$; $\sum_{i=1}^2p(A_i,B'_j)= \sum_{i=1}^2p(A'_i,B'_j)$, $j=1,2$.} Some authors consider that the violation of the inequalities is not conclusive of the presence of entanglement, if the marginal law is also violated \cite{dk2014,dkczj2016}. Without going into the specifics of this issue, we observe the following. The data we presented in this article do violate the marginal law, although not in a very pronounced way. This does not mean, however, that they cannot be modeled in a quantum theoretical way. For this, it is sufficent to introduce `entangled measurements' in addition to `entangled states', as was done in the modeling of previous cognitive experiments, also revealing the presence of entanglement in human cognition \cite{as2014}. But more importantly, it is possible to show that the violation of the marginal law in our experiment is not fundamental, but accidental, therefore in no way indicative of the fact that one cannot conclude about the presence of genuine entanglement effects revealed by our data. 

Indeed, measurement $A$ was taken to correspond to the {\it South--North} axis, with measurement $B$ then rotated of an angle of $45^\circ$ clockwise with respect to $A$, corresponding to the {\it Southwest--Northeast} axis, and measurements $A'$ and $B'$ rotated of an angle of $90^\circ$ clockwise with respect to $A$ and $B$, corresponding to the {\it East--West} and {\it Southeast--Northwest} axes, respectively. This specific choice of axis for the $A$-measurement (and consequently for the other rotated measurements) introduced a distinction (a first symmetry breaking) between the concept of `spatial direction' and the more specific concept of `wind direction'. However, a symmetrized version of the {\it Two Different Wind Directions} experiment can be designed, whose data will then obey the marginal law but still violate the CHSH inequality with same magnitude. This will be explained in detail in the second part of our article \cite{ass2017}, where an explicit quantum modeling in Hilbert space of the experimental data will also be given, using a singlet state to describe the initial state of the {\it Two Different Wind Directions} conceptual entity, and product measurements to describe the four different joint measurements that are executed on it.

\end{document}